# Pseudo Random Number Generator using Internet-of-Things Techniques on Portable Field-Programmable-Gate-Array Platform

Tee Hui Teo[1], Xinlong Zhang[2], Guangming Ren[2] and Chiang Liang Kok[2]

**Abstract**
This paper conducts a comparative study of three IoT-based PRNG models, including Logistic Map, Double Pendulum, and Multi-LFSR, implemented on an FPGA platform. Comparisons are made across key performance metrics like randomness, latency, power consumption, hardware resource usage, energy efficiency, scalability, and application suitability. Compared to Multi-LFSR, Logistic Map, and Double Pendulum Models provide perfect quality randomness, which is quite apt for high-security grade applications; however, the requirements of these models concerning power and hardware resources are also considerably high. By contrast, the Multi-LFSR comes into its own due to its lower latency, power consumption, and resource-efficient design. It is, therefore, suited for embedded or real-time applications. Furthermore, environmental sensors will also be introduced as entropy sources for the PRNGs to enhance the randomness of the systems, particularly in IoT-enabled battery-powered FPGA platforms. The experimental results confirm that the Multi-LFSR model has the highest energy efficiency, while the Logistic Map and Double Pendulum outperform in generating numbers with very high security. The study thus provides a deeper insight into decision-making for selecting PRNG models.

**Keywords**
Pseudo-Random Number Generator, Logistic Map, Double Pendulum, Multi-LFSR, FPGA, Energy Efficiency, Internet of Things, Real-Time Applications, Embedded Systems, Hardware-Based Randomness, High-Security Applications, Environmental Sensors, Portable FPGA Platform.

## Introduction

Algorithmic random number generators are everywhere, from simulation to computational creativity, and are used for all kinds of tasks. One of the generators is a Pseudo-Random Number Generator (PRNG). Chaotic PRNGs, such as those based on spatial chaotic maps of the Logistic type, have been shown to improve complexity and sensitivity, making them ideal for Monte Carlo simulations (Al-Daraiseh et al.2023). Hardware-based PRNGs using multiple Linear Feedback Shift Register (LFSRs) have enhanced simulation capabilities, offering scalability and real-time performance for secure environments (Akter et al. 2024). Moreover, PRNGs utilizing two-dimensional (2D) chaotic mappings provide improved randomness and uniformity, critical for cryptography and simulation applications (Hu2023).

PRNGs play a big part in machine learning frameworks like TensorFlow and PyTorch to produce stochastic streams for optimization and regularization techniques, such as Stochastic Gradient Descent (SGD) and dropout. However, research highlights that commonly used algorithms like Mersenne Twister (Matsumoto and Nishimura 1998; Marsland 2014), and Permuted Congruential Generator (PCG), (Bouillaguet et al.2020) allow for great randomness; alongside this, another computational efficiency problem. This problem led to inconsistent results in the outcome across different platforms (Antunes and Hill 2024). Furthermore, the choice of seed values has introduced variability in results, raising concerns about the robustness of empirical findings and the potential for "p-hacking" in sensitive applications (Naimi et al.2024).

Recently, by harnessing the power of Generative Adversarial Networks (GANs), PRNGs have received their conceptual design, permitting a variety of new possibilities for slight randomness generation. These GAN-based PRNGs generate statistically viable randomness with rigorous tests like acceptance from the National Institute of Standards and Technology (NIST) suite, rendering them suitable for data creation and cryptographic purposes (Wu et al. 2024). Similarly, Wasserstein GAN (WGAN), (Weng2019) methods remove issues like overfitting and ensure high-quality randomness for machine learning applications (Okada et al.2023). A newer approach is combining chaotic generators with numerical solvers. This way, given chaos's capability to generate dense pseudo-random sequences, the method provides greater security and unpredictability, especially for cryptographic and high-security machine-learning applications (Yacoub et al.2023).

[1]Singapore University of Technology and Design, Singapore;
[2]The University of Newcastle, Australia

**Corresponding author:**
Tee Hui Teo, Singapore University of Technology and Design,
8 Somapah Road, Singapore 487372.
Email: tthui@sutd.edu.sg

This paper proposes a field-programmable gate array (FPGA) implementation of a pseudo-random number generator (PRNG) built by Internet-of-Things (IoT) techniques. The PRNG is fed by multiple chaotic models and sensor-based entropy sources to provide increased randomness. The FPGA-based architecture can include real-time IoT data acquisition, processing, and communication distribution, making it particularly applicable for portable security solutions.

## Review of Hardware-based Pseudo-Random Number Generators

PRNGs are usually implemented on different classes of integrated circuits or FPGAs due to their fast processing at the required flexibility in specific demanding applications. Chaotic systems, which include items such as the Lorenz attractors (Pal and Mukhopadhyay 2020) and elementary cellular automata (Du et al. 2024), also find standard hardware implementations, because such systems have an intrinsic ability to produce pseudo-random sequences. Hardware implementations of PRNGs based on chaotic systems were then joined with a block-cipher system to form random encryption keys, effectively solving both the speed requirements and safety in encryption and decryption of digital images (Akter et al. 2024). When realized on hardware, these chaotic systems demonstrate excellent throughput, with some systems achieving speeds upwards of 49,946.62 Mbps (Abbassi et al. 2022). Additionally, integrating advanced chaotic systems, like the Half-Unit-Biased (HUB) format, into PRNG designs has improved their efficiency and chaotic behavior. This integration ensures the generated numbers maintain unpredictability while minimizing resource consumption, a crucial factor in large-scale hardware applications (Anurag et al. 2024). Furthermore, a new hardware architecture for chaos-based PRNGs has been developed using the HUB format. This architecture enhances chaotic-based systems' performance and resource efficiency by combining tent and Bernoulli maps, ensuring assertive chaotic behavior and better pseudo-randomness. This development has significantly improved the speed and security of hardware PRNGs (Da Silva et al. 2023). These implementations often exhibit low resource utilization, making them suitable for embedded systems and IoT devices that require compact and energy-efficient designs (Gafsi et al.2023).

Research into hardware PRNGs is instead offering exciting avenues by exploiting novel transformation algorithms, like Multiple Deep-Dynamic Transformation (MDDT), for throughput enhancement. Throughput speeds of newer hardware designs are as high as 14.4 Gbps, making them ideal for higher-speed applications requiring secure random number generation, ranging from secure communication systems to simulations(Li et al. 2024). Finally, implementing strong PRNGs for encryption tasks, such as color image encryption, shows how these technologies can be adapted to different domains, providing fast and secure key generation for image-based cryptographic systems (Akter et al.2024).

Another prominent trend in hardware PRNGs is using Cryptographically Secure Pseudo-Random Number Generators (CSPRNGs). These systems are designed to be secure against various attacks, and they typically use cryptographic hash functions such as Secure Hash Algorithm-2 (SHA-2) or SHA-3. Recent research has highlighted the advantages of using the SHA-3 algorithm for hardware-based CSPRNGs, noting improvements in energy efficiency and performance and better resource utilization compared to SHA-2-based solutions (Crocetti2023). These improvements make SHA-3 an attractive choice for designing secure PRNGs in hardware, especially in applications requiring a high degree of security, such as blockchain and secure communications (Gafsi et al. 2023). Even though hardware PRNGs have progressed enormously, there still exist challenges. One of the core issues is ensuring that the randomness not only appears random statistically but can also resist physical attacks. Hardware-based attacks like side-channel analysis can sometimes exploit weaknesses in the PRNG's design. Researchers have been looking at possible countermeasures to address such risks, including using noise sources, secured key management, and blended systems that combine the strengths of various extraction methods for random values.

## PRNG using FPGA

When designed on an FPGA, these PRNGs exhibit improved random bit sequence lengths, offering over 200 times better performance than simpler LFSR-based systems, which makes them suitable for high-performance systems (Akter et al. 2024). The use of chaotic systems for FPGA-based PRNGs has been explored in various studies. One approach uses the system's chaotic dynamics to produce fast and high-quality random numbers based on an n-dimensional non-degenerate chaotic system, which was implemented (Luo et al. 2024). The alternative method is to use multiple deep-dynamic transformations to increase the entropy and throughput of PRNGs. The technique focuses on how different iterative transformations can be used to suitably address the problem of improving the quality of randomness while at the same time making appropriate use of the available FPGA space. The approach was observed to surpass traditional designs of PRNGs in terms of speed and randomness, endowing it with a greatly optimized solution for high-performance applications (Li et al.2024). Another development is the introduction of lightweight reseeding schemes for FPGA-based PRNGs. Reseeding refreshes the internal state of the PRNG, preventing the generator from becoming predictable over time (Sivaraman et al.2024). This technique is beneficial in environments where continuous randomness is required, such as in wireless sensor networks, where energy efficiency and low resource consumption are essential. The reseeding mechanism ensures that the PRNG remains robust and unpredictable throughout its operation,thus enhancing its suitability for long-term use in wireless applications. Similarly, integrating Artificial Neural Networks (ANNs) with chaotic systems for PRNGs has been studied to enhance entropy generation further. Lastly, ANNs are combined with chaotic systems to optimize the randomness of the generated sequences (Sharobim et al. 2023). The ability of ANNs to learn and adapt to complex patterns is utilized to fine-tune the chaotic system's behavior, ensuring the production of high-entropy random numbers.

# Design of PRNG using IoT Techniques in Portable FPGA Platforms

Logistic map-based PRNG, Double Pendulum-based PRNG, and Multi-LFSR-based PRNG were implemented on the Xilinx CMOD-A7 FPGA, utilizing its robust processing capabilities to generate high-quality random numbers in real-time.

## *The Logistic Map Based Pseudo-Random Number Generators for FPGA Implementations*

Equation (1) defines the Logistic Map algorithm.

$$x_{n+1} = r \cdot x_n(1 - x_n), \quad (1)$$

It serves as a base model for chaotic behaviors, demonstrating an aspect of determinism behind what appears to be complicated and uncorrelated behavior. The behavior of r is crucial in this case, given that it dictates when stable behavior gives way to chaotic behavior. Here, between 1 and 3 inclusive means fixed point attractors, while beyond 3.57, this leads to chaotic dynamics, in which very slight changes in initial conditions lead to vastly different outcomes (Teo et al. 2024a). Several studies have highlighted the logistic map's potential for generating high-quality random numbers. For example, in cryptographic applications, the chaotic nature of the logistic map enables the production of unpredictable sequences, crucial for secure encryption algorithms. The logistic map's recursive and iterative structure also makes it suitable for hardware implementations, particularly on FPGAs. There, it can be integrated with other chaotic systems to enhance both randomness and performance. Additionally, the logistic map's applicability extends beyond cryptography into population modeling and financial forecasting, where its chaotic behavior can mimic real-world dynamic systems. Using a Logistic Map algorithm, an FPGA-based PRNG system enhances randomness by leveraging its chaotic properties. The system utilizes a Central Limit Theorem (CLT) transformation to refine the distribution of generated numbers into a Gaussian form, improving statistical properties (Teo et al. 2024a).

## *LFSR-Based Pseudo-Random Number Generators for FPGA Implementations*

Today, LFSR-based pseudo-random number generators are two popular techniques implemented in FPGA designs mainly because of their efficiency, simplicity, and low hardware resource requirements. For LFSR, bits are shifted through a register by applying a deterministic feedback function based typically on an XOR structure, which utilizes algebraic structures known as primitive polynomials to ensure maximum length of the sequence generated by the LFSR function and enhanced quality of randomness that is obtainable throughout the construction of any design function. Let's then write the generic LFSR state transition equations (2):

$$S_{t+1} = (S_t \gg 1) \oplus (S_t \cdot P) \quad (2)$$

where:

- $S_t$ represents the current state of the LFSR,
- $S_{t+1}$ is the next state after shifting,
- $\gg 1$ denotes a right shift by one position,
- $P$ is the primitive polynomial defining the feedback taps,
- $\oplus$ represents the XOR operation.

A commonly used maximal-length LFSR can be described by the equation 3:

$$X_n = (X_{n-k} \oplus X_{n-m}) \quad (3)$$

Where k, m are predefined tap positions chosen based on the selected primitive polynomial.

A dual-LFSR approach was proposed, integrating two LFSRs with an XOR gate to enhance randomness while maintaining low resource consumption on a Basys3 FPGA. The design significantly improved the sequence length by over 200 times compared to traditional single LFSR-based PRNGs, demonstrating the advantages of using multiple polynomials for enhanced entropy (Akter et al. 2024). Similarly, a study on low-cost FPGA implementations for Quantum Key Distribution (QKD) applications explored non-linearity integration in LFSR-based PRNGs, ensuring better unpredictability (Chandravanshi et al. 2023). The research highlighted that minimal non-linearity could increase randomness while maintaining computational efficiency. Further improvements in multi-bit LFSR architectures were examined where different primitive polynomials were analyzed for their impact on randomness quality and hardware performance (Sony et al. 2022). FPGA synthesis and implementation results revealed that selecting optimal polynomials affects the statistical properties of generated sequences. Meanwhile, a reversible LFSR design was proposed, aiming to reduce power consumption by 10% compared to traditional irreversible approaches. The study evaluated 4, 8, 16, and 32-bit LFSRs on FPGA platforms, demonstrating that the power efficiency of PRNGs can be improved without compromising randomness quality (Bailey et al. 2022). A Multi-LFSR PRNG enhanced with IoT sensor inputs to improve entropy and unpredictability. Combining multiple LFSRs with different feedback polynomials overcomes periodicity issues and generates high-quality random sequences. Environmental sensors provide additional randomness, ensuring robust security (Teo et al. 2024b). These studies collectively illustrate the adaptability of LFSR-based PRNGs for various FPGA applications, reinforcing the importance of feedback polynomial selection and non-linearity enhancements in PRNG design.

## *Double Pendulum*

Due to its highly nonlinear nature, the double pendulum system has become an essential object of study in chaotic dynamics. Its governing equations are typically derived using the Lagrangian formulation, combining the system's kinetic energy equation 5 and potential energy equation 6 to obtain a system of nonlinear coupled differential equations (4), (Kumar et al. 2019).

$$\frac{d}{dt}\left(\frac{\partial L}{\partial \dot{\theta}_i}\right) - \frac{\partial L}{\partial \theta_i} = 0, \quad i = 1, 2 \tag{4}$$

$$L = KE - U$$

where KE is the kinetic energy and U is the potential energy.

$$KE = \frac{1}{2}m_1 l_1^2 \dot{\theta}_1^2 + \frac{1}{2}m_2 [(l_1 \dot{\theta}_1)^2 + (l_2 \dot{\theta}_2)^2 + 2l_1 l_2 \dot{\theta}_1 \dot{\theta}_2 \cos(\theta_1 - \theta_2)] \tag{5}$$

$$U = -m_1 g l_1 \cos\theta_1 - m_2 g(l_1 \cos\theta_1 + l_2 \cos\theta_2) \tag{6}$$

The pendulum's starting position, lengths, and weights comprise these basic conditions. They serve as the random number generator's seed. These values are then applied to the governing equations (7) & (8) that describe the double pendulum. A system employs an FPGA to compute the chaotic motion equations of a double pendulum, and the generated numbers are analyzed through histogram and time-series tests, confirming strong randomness properties, (Teo et al.2024c).

$$d_1 = L_1(2m_1 + m_2 - m_2 \cos(2\theta_1 - 2\theta_2)) \tag{7}$$

$$d_2 = L_2(2m_1 + m_2 - m_2 \cos(2\theta_1 - 2\theta_2)) \tag{8}$$

The angular acceleration equations (9) & (10) are derived as:

$$\omega_1' = \frac{1}{d_1}\Big(-g(2m_1 + m_2)\sin\theta_1 - m_2 g \sin(\theta_1 - 2\theta_2) - 2\sin(\theta_1 - \theta_2)m_2 \left[\omega_2^2 L_2 + \omega_1^2 L_1 \cos(\theta_1 - \theta_2)\right]\Big) \tag{9}$$

$$\omega_2' = \frac{2\sin(\theta_1 - \theta_2)}{d_2}\Big(\omega_1^2 L_1(m_1 + m_2) + g(m_1 + m_2)\cos\theta_1 + \omega_2^2 L_2 m_2 \cos(\theta_1 - \theta_2)\Big) \tag{10}$$

Where:

- $\omega_1 = \dot{\theta}_1$, $\omega_2 = \dot{\theta}_2$ (angular velocity)
- θ : angle of pendulum (0 is vertical downwards, counterclockwise is positive)
- L : length of the rod (constant)
- T : tension in the rod
- m : mass of the pendulum
- g : gravitational constant

Recent research has focused on analyzing the role of saddle points in the global phase space of the double pendulum's chaotic motion (Kaheman et al.2023). Studies show that saddle points form boundaries near unstable periodic orbits, causing complex trajectory transitions in the chaotic phase space (Levien and Tan 1993). Experimental validations have further supported the accuracy of numerical simulations (Maiti et al. 2016). Furthermore, bifurcation analysis has revealed the effects of different initial conditions on chaotic behavior and introduced methods to measure the intensity of chaos based on basins of attraction (Qin and Zhang2024).

The straightforward construction of the Double Pendulum system, but with the highly complex chaotic features, makes it a popular system for validating chaos theory and nonlinear dynamics methods. The research underway ranges from purely experimental and numerical studies of chaotic features of the system, to more applied work in physics related to control systems and aerospace dynamics in connection with saddle point transport theory in orbital dynamics (Kaheman et al. 2023). Future research could further integrate machine learning techniques to predict chaotic trajectories and optimize control strategies to stabilize chaotic motion (Rafat et al.2009).

**Results**

Table 1 shows the power consumption of the three PRNG models. It can be observed that the Logistic Map requires the highest power consumption in the logic and signal processing stages, mainly due to the complexity of the chaotic system. On the other hand, the Double Pendulum and Multi-LFSR models exhibit lower power consumption, making them more energy-efficient in the signal and logic stages. The Double Pendulum model has lower power consumption than the Logistic Map, which makes it a more suitable candidate for low-power applications.

|         | Logistic Map | Double Pendulum | Multi-LFSR |
|---------|--------------|-----------------|------------|
| Static  | 0.485W       | 0.075W          | 0.077W     |
| Signals | 8.762W       | 0.546W          | 0.787W     |
| Logic   | 10.047W      | 0.443W          | 0.491W     |
| I/O     | 7.629W       | -               | -          |
| DSP     | 2.629W       | 0.812W          | 1.069W     |

**Table 1.** The Power Consumption of three PRNG models—Logistic Map, Double Pendulum, and Multi-LFSR

Table 2 summarizes the resource utilization in terms. The Logistic Map requires significantly more LUTs and flip-flops than the other models, indicating its complexity in hardware implementation. The Double Pendulum and Multi-LFSR models are more resource-efficient, requiring fewer resources for the same functionality. The Multi-LFSR is the most resource-efficient model, requiring the least LUTs, flip-flops, and I/Opins, making it a good candidate for embedded applications where resource constraints are a concern.

|     | Logistic Map | Double Pendulum | Multi-LFSR |
|-----|--------------|-----------------|------------|
| LUT | 663          | 126             | 71         |
| FF  | 169          | 143             | 68         |
| DSP | 3            | -               | -          |
| I/O | 22           | 15              | 34         |

**Table 2.** The Resource Utilization of three PRNG models—Logistic Map, Double Pendulum, and Multi-LFSR

To determine the efficacy of three different designs of pseudo-random number generators (Multi-LFSR, Logistic

|  | Logistic Map | Double Pendulum | Multi-LFSR |
|---|---|---|---|
| Randomness | High | High | Medium |
| Latency | High | Medium | Low |
| Power Consumption | 29.559W(98% Dynamic) | 1.877W (96% Dynamic) | 2.451W (97% Dynamic) |
| Hardware Resource Usage | High | Medium | Low |
| Energy Efficiency | Low | High | Very High |
| Scalability | Low | Medium | High |
| Application Suitability | High-Security | High-Security | Embedded, Real-time |

**Table 3.** Comparison of three PRNG models—Logistic Map, Double Pendulum, and Multi-LFSR

Map, and Double Pendulum), we measured their Latency on FPGA and compared them for various applications. The FPGA is an Xilinx CMOD-A7 that utilizes clock cycle counting, ILA, and UART transmission time measurement to evaluate how long it takes to generate random numbers with each algorithm. We found that the fastest was the Multi-LFSR PRNG, requiring only 1-2 clock cycles (10-20 ns), due to the simple shift register used, with potential applications in high-speed encryption where very low latency is a factor. On the other hand, the Logistic Map PRNG, being nonlinear and chaotic in computations, required an average of 5-10 clock cycles (50-100 ns), thus making it applicable to most embedded systems with reasonable randomness. On the other hand, the Double Pendulum PRNG, whose time latency is longer due to higher computing needs like trigonometric operations and, hence, takes around 20- 50+ clock cycles (200-500 ns), meaning it can be used for any applications requiring high security where real-time constraints are not minimized.

In Table 3. On the one hand, the Logistic Map implies high power consumption (29.559W) and low utilization of hardware resources. Thus, it is the least energy-efficient. By contrast, the Double Pendulum has maximum power efficiency (1.877W) and a medium-exhaustion of hardware resources, but still performs poorly as the Multi-LFSR. The Multi-LFSR is well-known for its low power consumption (2.451W) and lean resource utilization. Thus, it is well-suited for embedded and real-time applications, where low latency and high energy efficiency are demanded. Additionally, it will scale very well with extra resources to accommodate larger systems. While the Logistic Map and Double Pendulum are applicable for applications that need very high randomness and security, the Multi-LFSR is credited when resource limitations and real-time performance are the main priority.

## Conclusion

This research paper has endeavored to thoroughly compare three hardware-based pseudo-random number generator models, namely Logistic Map, Double Pendulum, and Multi-LFSR, as embodied in the Xilinx CMOD-A7 FPGA. The analysis states that Logistic Map and Double Pendulum models, which offer very high randomness and are therefore aligned with high-security applications, consume more power and require more hardware resources. On the other hand, the Multi-LFSR model has less latency, lower power consumption, and more efficient usage of FPGA resources, which appears to be most suitable for embedded systems and real-time applications, in which low energy consumption and performance in a short time are essential. Furthermore, incorporating environmental sensors as entropy sources creates improved randomness, bringing additional flexibility to the PRNGs, which are suitable for IoT-enabled, battery-powered FPGA platforms. The model of the Multi-LFSR is therefore rated higher in terms of energy use, scale-up mechanism development, and further application portability. In contrast, the Logistic Map and Double Pendulum models would fit into the applications requiring more security, although at a computational cost. Therefore, the final choice should be made according to the specific application requirement regarding the degree of randomness, security, resource efficiency, or performance within a real-time setting. Future works will be focused on optimizing such models for ASIC implementation designs to enhance energy efficiency and adaptive entropy techniques in further vulnerability against challenges for high-complexity hardware-based random number generators in IoT surroundings.

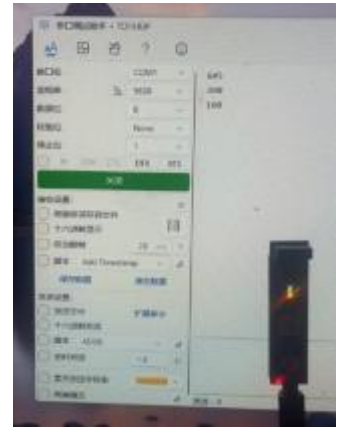

**Figure 1.** Real-time Pseudo Random Number Generator Display